\documentclass[preprint,byrevtex,showpacs, prb]{revtex4}
\usepackage{graphicx}% Include figure files
\usepackage{dcolumn}% Align table columns on decimal point
\usepackage{bm}% bold math
\usepackage{epsfig}
\usepackage{times}
\usepackage{amsmath, amsthm}
\usepackage{amssymb}
\usepackage{textcomp}
\begin{document}
\title{Optical switching in graded plasmonic waveguides}
\author{J. J. Xiao}
\affiliation {Department of Physics, The Chinese University of Hong Kong,
Shatin, New Territories, Hong Kong, China}
\author{K. Yakubo}
\affiliation{Division of Applied Physics, Graduate School of Engineering,
Hokkaido University, N13-W8, Sapporo 060-8628, Japan}
\author{K. W. Yu\footnote{Electronic address: kwyu@phy.cuhk.edu.hk}}
\affiliation{Department of Physics and Institute of Theoretical Physics, The
Chinese University of Hong Kong, Shatin, New Territories, Hong Kong, China}

\begin{abstract}

A new mechanism of longitudinal confinement of optical energy via coupled plasmon
modes is proposed in chains of noble metal nanoparticles embedded in a graded
dielectric medium, which is analogous to the confinement of electrons in
semiconductor quantum wells. In these systems, one can control the
transmission of optical energy by varying the graded refractive index of the
host medium or the separation between the nanoparticles to realize the
photonic analogue of electronic transistors. Possible passband tunability by
nanoparticle spacing and modulation of the refractive index in the host
medium have been presented explicitly and compared favorably with numerical
calculations.
\end{abstract}
\date{\today}

\maketitle

One of the most attractive studies in advanced optics is to conquer the
general difficulties in conventional photonics: (1) restriction of the
diffraction limit, which prevents achieving high degree of miniaturization
and monolithic integration of optical devices and circuits; and (2) light
fields existing in three dimensions, inhibiting integrated on-chip devices. A
prominent way to overcome these two major obstacles is related to use of
surface plasmons (SPs) in plasmonic structures, e.g., see Refs.~1-8
%\cite{Shalaev05, Zayats03,Barnes03, Hutter04,Girard05,Zayats05, Atwater05}
and references cited therein. Along with the advancement of nanofabrication
techniques, many nanostructures and devices based on SPs have been studied,
for example, metallic nanoparticles, \cite{Hutter04} nanowires,
\cite{Shalaev05} nanostrips, \cite{Barnes03} and metal grooves
\cite{Groove04} and wedges, \cite{Wedges05} as well as integrated components
such as nanoparticle chains, \cite{Atwater05}  multilayers, \cite{Overfelt05}
and combined plasmonic and dielectric waveguides.
\cite{IntegraedPlasmonDielectric} They essentially utilize SPs features of
\textit{lateral/transverse} localization and evanescence, \cite{Karalis05}
and their operations in near-field regime, thus can easily beat the above
mentioned restrictions in conventional photonics.

Although these plasmonic systems offer the advantages to relieve the
difficulties in the miniaturization and integration of photonic devices, SPs
suffer from intrinsic dissipation of consisted material, e.g., mostly noble
metals like gold and silver and thus decay excessively along the propagation
direction, or referred to the longitudinal direction. Nevertheless, their
propagation length is still satisfactory in nanooptics, e.g.,  $1/e$
propagation lengths of a few tens of $\mu$m are reported. \cite{longSPP} It
is, alternatively, desirable to have localized SPs in the longitudinal
direction, in addition to their naturally lateral/transverse confinement. In
this sense, they may sustain modes in ultrasmall volume and highly confined
electromagnetic field. To this end, only very few works have focused on this
issue, for instance, nano-concentration of optical energy by localizing SPs
in graded plasmonic waveguides (i.e., axially nonuniform waveguides),
\cite{Stockman04} and symmetry breaking in plasmonic chains, which helps the
enhancement of local fields, and facilitates the forming of localization of
SPs due to asymmetric polarization charge distribution. \cite{Bergman03}
Other examples include using nanoscale Fabry-P\'{e}rot cavity to confine SPs
inside, \cite{FabryPerot} and constructing plasmonic crystals by decorating
or structuring metallic surfaces at nanoscale. \cite{Zayats03,Baumberg05} In
analogy of photonic crystals, coupled SPs show localization near the
band-edges in these plasmonic crystals and demonstrate desirable
localizations.

In this work, we show that precise control of \textit{longitudinal
localization} of SPs in plasmonic nanoparticle chains (PNCs) is possible by
simply imposing a gradient in the refractive indices of the surrounding host
medium, which can behave like a plasmonic switcher. The localization
mechanism is different from those of the two extreme cases: periodically
structured plasmonic crystals \cite{Zayats03,Stroud04,Atwater00} and
disordered plasmonic systems. \cite{Bergman01,Shalaev05PRB} In fact,
closely-separated PNCs were proposed for potential applications in energy
guiding and routing inside the subwavelength scale. \cite{Atwater05} Also
they were demonstrated efficiently converting propagating light (far-fields)
to SPs (near-fields).\cite{Ohtsu05} The precise control of SPs in our
proposed system can also achieve confinement of electromagnetic fields in
specific region to control their interactions with specimen. \cite{Quidant05}
Furthermore, we can envision applications of such PNC arrays that
evanescently couple two dielectric slab waveguides, e.g., in
photo-addressable birefringent polymer matrix  (optical control) or in
electro-optical materials such as liquid crystals (electrical control).

Let us assume that a chain of $N$ identical metallic nanoparticles of Drude
form dielectric function $\epsilon_m(\omega)=1-\omega_p^2/\omega^2$
(lossless) is immersed in a dielectric host with its dielectric function
varying from the left-hand side to the right-hand side along $x$-axis as
\begin{equation}
\label{eq:gradhost} \epsilon_2(x)=\epsilon_{\text{left}}+cx/l,
\end{equation}
where $x$ denotes the position of the particles, $l$ the total length of the
chain, and $c$ the coefficient of dielectric gradient which can be formed
naturally, induced by a pump light that possesses intensity variation
spatially, or by possible ultrasonic induced gradient in a piezoelectic
host.\cite{Silly05} The $i$-th and the $j$-th nanoparticles are separated by
a distance $d_{ij}=d_0|i-j|$, with their position denoted by $x_i=d_0(i-1)$,
where $i=1, 2, \cdots, N$. This means that the nanoparticles are equally
spaced in a periodical chain which is immersed in a host with dielectric
function increasing from $\epsilon_{\text{left}}$ at the chain's leftmost
site to $\epsilon_{\text{left}}+c$ at the rightmost site.
%rather than gradually spaced.
Therefore, each spherical nanoparticle acquires an isotropic dipole
polarizability
\begin{equation}
\label{eq:dipolefactor}
\beta_i=\epsilon_2(x_i)a^3\frac{\epsilon_m(\omega)-\epsilon_2(x_i)}
{\epsilon_m(\omega)+2\epsilon_2(x_i)}.
\end{equation}
The dipolar resonant frequency is denoted by
$\omega_i=\sqrt{\omega_p^2/[1+2\epsilon_2(x_i)]}$ which makes the real part
of denominator of Eq.~\eqref{eq:dipolefactor} zero. \cite{Shalaev05} In the
coupled dipole approximation, the self-consistent coupling equations of the
local fields ${\bf E}_i$ around the $i$-th nanoparticle read, \cite{Girard05}
\begin{eqnarray}
 {\bf E}_i=\sum_{j \neq i}^N {\tilde{\boldsymbol{T}}}{(i, j)} \cdot
(\beta_j {\bf E}_j)+{\bf E}_i^{(0)}, ~ (i=1, 2, \cdots N),
\label{eq:localfield}
\end{eqnarray}
where ${\bf E}_i^{(0)}$ represents the external electric field, and the
dipole coupling is given by $\tilde {\boldsymbol
T}(i,j)={\boldsymbol{T}}(i,j)/\epsilon_2(x_i)$ where ${\boldsymbol{T}}(i,j)$
denotes the near field coupling in vacuum. In Cartesian coordinates ($x$,
$y$, $z$), the ($\beta$,$\gamma$) components are given by
\begin{equation}
\boldsymbol{T}_{\beta,\gamma}(i, j)=\frac{3d_{ij, \beta} d_{ij, \gamma}
-|{\bf d}_{ij}|^2 \delta_{\beta,\gamma}}{|{\bf d}_{ij}|^5}.
\end{equation}
Where ${\bf d}_{ij}$ represents the vectorial distance with $\beta$ or
$\gamma$ component $d_{ij, \beta|\gamma}$. After substituting the Drude form
into Eq.~\eqref{eq:dipolefactor} and eliminating the bulk plasmon frequency
$\omega_p$ in Eq.~\eqref{eq:localfield}, we have the following coupled
equation of motion for the dipole moments
\begin{equation}
\label{eq:eig} -\omega^2 p_i=-\omega_i^2p_i+\frac{3\lambda_{(L,T)}\omega_i^2
\epsilon_2^2(x_i)}{1+2\epsilon_2(x_i)} \Big[E^{(0)}_{i}+\sum\limits_{j \neq
i}^N \left(\frac{a}{d_{ij}}\right)^3\frac{p_j}{\epsilon_2(x_i)}\Big],
\end{equation}
where $\lambda_{\text{L}}=2$ and $\lambda_{\text{T}}=-1$ are for the cases of
longitudinally (i.e., parallel to the chain axis) and transversely polarized
${\bf E}_i^{(0)}$ (simply indicated by $E_i^{(0)}$), respectively. In view of
the fact that we are concerned with the plasmon waves near the resonant
frequency,\cite{Atwater00,Stroud04} i.e., $\omega \sim \omega_i$, we have
obtained Eq.~\eqref{eq:eig} with the approximation
$[\epsilon_2(x_i)-1]\omega^2+[2\epsilon_2(x_i)+1]\omega_i^2 \approx
3\epsilon_2(x_i)\omega_i^2$. Equation \eqref{eq:eig} captures the main
features of the plasmonic coupling between nanoparticles in the PNCs. For
example, it shows that the coupling depends on the resonant frequencies, the
polarization, and the properties of the metal and the surrounding host, as
well as the separations between the particles. \cite{Atwater05}

We seek plasmonic eigenmode solutions $\Phi_{\alpha}(i)$ of
Eq.~\eqref{eq:eig} in the absence of the external driving term $E^{(0)}_{i}$.
The results are shown in Fig.~\ref{fig:fig1} for the longitudinal
polarization and with parameters listed below: $\epsilon_{\text{left}}=3.0$,
$c=1.0$, $N=100$, and $d_0=3a$, namely $l=3a(N-1)$, where $a$ is the radii of
the nanoparticles. In this nanoparticle separation, complications from
multipolar interactions are negligible \cite{Xiao05} and our approximations
are quite accurate. It is noteworthy that in the calculations we used the
approximation that the dielectric constant of the host around the $i$-th
nanoparticle is homogeneous, e.g., denoted by $\epsilon_2(x_i)$. In this
case, we actually have a host dielectric function $\epsilon_2(x_i)$ gradually
varying from $3.0$ at the left-hand extremity to $4.0$ at the right-hand
extremity. The solid curve in Fig.~\ref{fig:fig1}(a) represents the
relationship (i.e., pseudo-dispersion relation) between the coupled plasmon
frequency $\omega$ and the mode index $\alpha$ for the longitudinal
polarization. Longitudinal results for the same PNC within homogenous hosts
of dielectric constant $\epsilon_2=3.0$ (dash-dotted line) and
$\epsilon_2=4.0$ (dashed line) are also shown. It is noticed that the
presence of a gradient in the host refractive index alters the plasmon
(pseudo-) dispersion relation dramatically. This is also signified in the
density of states (DOS)
$D(\omega)=\sum_{\alpha}\delta(\omega-\omega_{\alpha})/N$, as shown in
Fig.~\ref{fig:fig1}(b). In detail, the DOS peaks at band-edges of the
homogenous host cases (dashed and dash-dotted lines) are shifted toward the
Mie dipole plasmon resonant frequencies
$\omega_0=\omega_p/\sqrt{1+2\epsilon_2}$, due to the gradient in the host
refractive index. More specifically, for the graded host case we studied,
there are two peaks in the DOS at around $\omega_{\text{f1}}=0.331\omega_p$
and $\omega_{\text{f2}}=0.364\omega_p$, for near field full couplings (Full)
between nanoparticles. The two peaks, however, appear at
$\omega_{\text{c1}}=0.337\omega_p$ and $\omega_{\text{c2}}=0.367\omega_p$ if
only the nearest-neighboring couplings (NN) are taken into account. These two
peaks in fact correspond to two points of transition between confined modes
and extended modes, which is easily seen in Fig.~\ref{fig:fig1}(c) showing
the inverse participation ration (IPR) of the normal mode,\cite{gradon} as
well as directly from the mode patterns  (right panels in
Fig.~\ref{fig:fig1}). Interestingly, the modes outside the frequency range
$\omega_{\text{f1}}<\omega<\omega_{\text{f2}}$ are localized at the two
extremities of the PNC, i.e., modes with $\omega>\omega_{\text{f2}}$ (high
frequencies) are localized at the left-hand side while modes with
$\omega<\omega_{\text{f1}}$ (low frequencies) are confined at the right-hand
side with dramatically increased IPRs. These localization/confinement are
clearly shown by the mode patterns in Figs.~\ref{fig:fig1}(d) and 1(f).
However, normal modes with frequency
$\omega_{\text{f1}}<\omega<\omega_{\text{f2}}$ are extended over the entire
PNC [see Fig.~\ref{fig:fig1}(e)]. They exhibit IPRs that are near constant as
shown in Fig.~\ref{fig:fig1}(c).

Intuitively, the existence of localized (confined) modes at the two ends is
reasonable in view of the fact that the Mie resonances is lower at the right
hand where there is a relatively larger refractive index in the host and
higher at the left hand side with a relatively smaller refractive index in
the host. But to further understand the localization-delocalization
transitions, let us neglect long range interactions and truncate the
summation in Eq.~\eqref{eq:eig} up to the nearest neighbors, i.e., consider
NN couplings. The resulting coupled plasmon system is similar to a chain of
graded coupled harmonic oscillators \cite{gradon} with additional on-site
harmonic potentials. In this case, the particle mass $m_{i}$, and the
strength of the additional harmonic spring $U_{i}$ are respectively
\begin{eqnarray}
m_{i}&=&\frac{1+2\epsilon_{2}(x_{i})}{3\lambda\omega_{i}^{2}
\epsilon_{2}(x_{i})}, ~~(i=1, 2, \cdots, N), \\
U_{i}&=&\frac{1+2\epsilon_{2}(x_{i})}{3\lambda \epsilon_{2}(x_{i})}-2K_{0},
~~(i=1, 2, \cdots, N),
\end{eqnarray}
where $K_{0}=(a/d_{0})^3$ is a force constant between adjacent particles that
depends on the ratio of the interparticle spacing and the particles radius.
Figures~\ref{fig:fig2}(a) and \ref{fig:fig2}(b) show $m_i$ and $U_i$ for the
case of $d_0=3a$ (solid lines) and $d_0=2.5a$ (dashed lines) respectively,
for $\lambda=\lambda_{\text{L}}=2$. In this case $m_{i}$ and $U_{i}$ are
respectively increasing and decreasing functions of $i$, and for
$\lambda=\lambda_{\text{T}}=-1$, they are of the same functions of $i$ (not
shown). Illustrating this situation, we can understand that the lower
frequency modes are confined at the right side of the chain and higher
frequency modes at the left side. The lowest frequency $\omega_{\text{L}}$
(i.e., lower bound) of extended modes is the lowest frequency of the
homogeneous system with $m_{i}=m_{1}$ and $U_{i}=U_{1}$ for any $i$. This
frequency $\omega_{\text{L}}$ is simply given by
\begin{equation}
\label{eq:min} \omega_{\text{L}}=\sqrt{U_{1}/m_{1}}.
\end{equation}
Also, the highest frequency $\omega_{\text{H}}$ (i.e., upper bound) of
extended modes is the highest frequency of the homogeneous system with
$m_{i}=m_{N}$ and $U_{i}=U_{N}$ for any $i$. This frequency is given by
\begin{equation}
\label{eq:max} \omega_{\text{H}}=\sqrt{\frac{4K_{0}+U_{N}}{m_{N}}}.
\end{equation}
For parameters $N=100$, $d_{0}=3a$, $c=1.0$, $\epsilon_{\text{left}}=3.0$,
and $\lambda_{\text{L}}=2$, Eqs.~\eqref{eq:min} and \eqref{eq:max} result in
$\omega_{\text{L}}=0.340\omega_p$ and $\omega_{\text{H}}=0.365\omega_p$,
respectively. These are in excellent agreement with the numerical data of
nearest-neighboring coupling as shown in Figs.~\ref{fig:fig1}(b) and 1(c),
where $\omega_{\text{L}}$ and $\omega_{\text{H}}$ are indicated by the two
vertical dashed lines. In Fig.~\ref{fig:fig1}(b),  they coincide respectively
with the two peaks in the DOS with NN couplings, which locate right at the
two transition frequencies of $\omega_{\text{c1}}=0.337\omega_p$ and
$\omega_{\text{c2}}=0.367\omega_p$.

The analytic results of the transition frequencies given by
Eqs.~\eqref{eq:min} and \eqref{eq:max} help to design the passband
substantially, although they are for nearest-neighboring coupling. In this
sense, we hereby attempt the tunable passband for the longitudinal
polarization represented by the analytical forms and confirm it by numerical
calculations. The results are demonstrated in Figs.~\ref{fig:fig2}(c) and
2(d), where we assume an additional nonlinear host dielectric variation,
i.e., $\epsilon_2(x)=(\epsilon_{\text{left}}+cx/l)(1+p)$. Here $p$ indicates
the change of dielectric constant due to optical nonlinear effects, i.e.,
$p=\chi E_{\text{pump}}^2$, where $\chi$ is the third order nonlinear
susceptibility of the host media and $E_{\text{pump}}$ represents a strong
pump field. In Fig.~\ref{fig:fig2}(c) we plot $\omega_{\text{L}}$ (solid
lines) and $\omega_{\text{H}}$ (dashed lines) as functions of $c$ with
$p=0.0$ for $d_0=3a$ and $d_0=2.5a$, respectively. We also show the results
of $\omega_{\text{c1}}$ (symbol $\square$) and $\omega_{\text{c2}}$ (symbol
$\bigcirc$) extracted from the numerical calculations. The numerical data
agree very well with the analytical ones. This figure indicates that with
fixed nanoparticle spacing $d_0$ in the PNC, $\omega_{\text{L}}$ keeps
relatively constant, while $\omega_{\text{H}}$ decreases almost linearly for
an increased gradient coefficient $c$ from, say, $0.2$ to $1.5$. On the other
hand, increasing the separation of the nanoparticles, e.g., changing
$d_0=2.5a$ to $d_0=3a$ narrows the passband dramatically, as both
$\omega_{\text{L}}$ and $\omega_{\text{H}}$ approach to the Mie resonant
frequency $\omega_i$ because the particles become more isolated, that is, the
couplings between them are diminished.

Figure~\ref{fig:fig2}(d) demonstrates tunability of the passband by
nonlinearity effect in the host medium. For example we show the transition
frequencies $\omega_{\text{L}}$ and $\omega_{\text{H}}$ versus $p$ for
$d_0=3a$ with two different gradient coefficient $c=0.5$ and $c=1.0$,
respectively. Similarly the numerical data of $\omega_{\text{c1}}$ (symbol
$\square$) and $\omega_{\text{c2}}$ (symbol $\bigcirc$) are compared
favorably to the analytical results (curves). It is seen that for increased
$p$, i.e., increased dielectric constant in the graded host, both the low and
high transition frequencies are red-shifted. These are ascribed to the fact
that the Mie resonant frequency is redshifted when the dielectric constant of
the surrounding host is increased. Interestingly, $\omega_{\text{L}}$
($\omega_{\text{c1}}$) decreases relatively at the same fashion as
$\omega_{\text{H}}$ ($\omega_{\text{c2}}$) does, varying over a span, e.g.,
from around $0.35\omega_p$ to a extremely low frequency. Admittedly, it is
hard to achieve such a large nonlinear modulation, for example, for $p>1.0$.
However, a relatively small modulation in the host dielectric nonlinearity,
e.g., from $p=0.0$ to $p=0.5$ still results a desirable tuning of
$\omega_{\text{L}}$ and $\omega_{\text{H}}$. We believe these results can
find potential applications of dynamical switching in plasmonics. The pump
field may also induce a nonlinearity in the metallic particles. However, the
nonlinearity in host is assumed to be larger, resulting in a graded
polarizability along the chain. Finally, the results for the transverse modes
($\lambda=-1$) are quite similar to those of the longitudinal modes
($\lambda=2$) presented here.

In summary, we propose the use of a gradient dielectric host to achieve the
precise confinement of electromagnetic fields in plasmonic nanoparticle
chains. It is demonstrated that the coupled surface plasmons in these systems
become easily tunable between confined modes and extended modes. We offer an
interpretation of the transition mechanism by mapping the problem onto a
graded coupled harmonic oscillators. In addition to possible operation for
plasmonic switching, these systems may be also of interest for applications
such as surface-enhanced Raman scattering or biosensing.

\hfill

This work was supported in part by the RGC Earmarked Grant of the Hong Kong
SAR Government (K.W.Y.), and in part by a Grant-in-Aid for Scientific
Research from Japan Society for the Promotion of Science (No.~16360044).

\newpage
\begin{center}
\textbf{Figure Captions}
\end{center}

\hfill

\begin{figure*}[htbp]\caption{(Color online) (a) Pseudo-dispersion relations
for full coupling. Those for cases of homogeneous host with $\epsilon_2=3.0$
(dash-dotted line) and $\epsilon_2=4.0$ (dashed line) can be regarded as the
usual dispersion relations. (b) Density of states (DOS) versus coupled
plasmon mode frequency for the full coupling. The line with circles represent
results of graded case for nearest-neighboring coupling. (c) Inverse
participation ratio (IPR) versus plasmon mode frequency. (d) Typical
excitation in low frequency regime ($\omega=0.315\omega_p$). (e) Typical
excitation near the resonant frequency ($\omega=0.353\omega_p$). (f) Typical
excitation  at high frequency ($\omega=0.390\omega_p$).}
 \label{fig:fig1}
\end{figure*}

\hfill

\begin{figure*}[htbp]\caption{(Color online) (a) Effective mass $m_i$ and
(b) on-site potential $U_i$ for $d_0=3a$ and $d_0=2.5a$ with $c=1.0$ and
$p=0.0$. (c) Analytical (curves) and numerical (symbols) transition
frequencies $\omega_{\text{L}}$ ($\omega_{\text{H}}$) and
$\omega_{\text{c1}}$ ($\omega_{\text{c2}}$) versus the host gradient
coefficient $c$ without nonlinearity ($p=0.0$) for different nanoparticle
distances $d_0$. (d) Analytical (curves) and numerical (symbols) transition
frequencies versus nonlinearity (pump field) modulation in the host medium,
with  $d_0=3a$ for different host gradient coefficient $c=0.5$ and $c=1.0$,
respectively.}
 \label{fig:fig2}
\end{figure*}

\newpage
\centerline{\includegraphics[scale=0.9]{Fig1.eps}} \centerline{Fig.1./Xiao,
Yakubo, and Yu}

\newpage
\centerline{\includegraphics[scale=0.6]{Fig2.eps}} \centerline{Fig.2./Xiao,
Yakubo, and Yu}

\end{document}